**An open-source, scalable and adaptable workflow for organizing echosounder data from transect surveys for machine learning applications**

Caesar Tuguinay[1,2], Wu-Jung Lee[1,a], Valentina Staneva[3], Elizabeth M. Phillips[4], Rebecca E. Thomas[4], Alicia Billings[4], Julia Clemons[4]

[1]Applied Physics Laboratory, University of Washington, Seattle, Washington, United States

[2]Department of Electrical and Computer Engineering, University of Washington, Seattle, Washington, United States

[3]eScience Institute, University of Washington, Seattle, Washington, United States

[4]Fisheries Engineering and Acoustic Technologies Team, Northwest Fisheries Science Center, National Marine Fisheries Service, National Oceanic and Atmospheric Administration, Seattle, Washington, United States

[a]Corresponding author: leewj@uw.edu

Keywords: data workflow, fisheries acoustics, echosounders, open-source software, machine learning, deep learning

Running head: Echosounder ML data workflow

# Abstract

Echosounders, or high-frequency active sonar systems, have become standard tools for quantifying and mapping the distribution of marine organisms in fisheries and ecological surveys. Conventional echosounder data analysis often relies on human annotation of echograms, which are sonar imagery formed by echo intensity. Over the past decade, in parallel with the exponential growth in the volume of echosounder data, there has been a corresponding increase in the development of machine learning (ML) methods that operate primarily on echograms as images. However, echograms are not simply images: they are associated with specific spatiotemporal coordinates that are essential for alignment with survey events, human annotations, and other oceanographic datasets. We present an adaptable two-stage workflow for constructing analysis-ready datasets for ML development tailored for echograms from transect surveys, in which (1) acoustic data are partitioned according to transect designation, and (2) masks are created from annotations referencing user-defined uniform spatiotemporal echo data grid. Importantly, ancillary information, such as geospatial coordinates and oceanographic measurements, is propagated across processing stages to preserve the essential contextual information for downstream analyses. We demonstrate the scalability of our workflow implementation based on two open-source software libraries, Echopype and Echoregions, using two example fisheries survey datasets. We additionally provide executable tutorials that guide readers through the computational implementation of this workflow. Together, these elements provide a scalable framework for creating analysis-ready echosounder datasets for ML applications.

# Introduction

Over the past decades, high-frequency active sonar systems, or echosounders, have been widely used to map the distribution of pelagic fish and zooplankton and to delineate seafloor topology (Medwin and Clay, 1998; Brown and Blondel, 2009; Stanton, 2012). These instruments operate by generating acoustic signals and capturing the returning echoes, which contain information about the scatterers, or objects that reflect the sound. The intensity of an echo is a function of the sound frequency and physical properties of the scatterer(s), and is often used in the echo interpretation process (Medwin and Clay, 1998; Simmonds and MacLennan, 2005; Jech and Michaels, 2006; Stanton, 2012).

To sustainably manage economically significant commercial fisheries, acoustic-trawl surveys are conducted to deliver estimates of fish biomass to support fishery stock assessments (Koslow, 2009; Greene et al., 2014). In these surveys, echosounders are typically mounted on the hull or a retractable keel of a vessel and directed downwards to acoustically profile the water column. As the echosounder transmits acoustic signals, or pings, along the vessel track, the intensity of the returning echoes can be aligned and color-coded to form two-dimensional echograms, in which each pixel is a digitized echo sample at a specific time and depth (top panel of Fig. 1C) (Maclennan, 2002).

Fisheries scientists use a variety of specialized software [e.g., Echoview (Echoview Software Pty Ltd, 2025), ESP3 (Ladroit et al., 2020), LSSS (Korneliussen et al., 2006), MOVIES3D (Trenkel et al., 2009), and Matecho (Perrot et al., 2018)] to perform both manual and automated echo data analysis. For automated analysis, many software packages provide rule-based algorithms or allow users to create their own routines to identify specific animal taxa, aggregation types, seafloor, noise, etc (e.g., Weill et al., 1993; Jech and Michaels, 2006; Kang et al., 2006; De Robertis and Higginbottom, 2007; Trenkel et al., 2009; Ryan et al., 2015; Perrot et al., 2018; Phillips et al., 2022). For example, age-0 walleye pollock off Hokkaido, Japan have frequency response characteristics that allow for rule-based identification and separation from other pollock age groups by comparing differences in echo strengths received at two frequencies (Kang et al., 2006). However, many animals exhibit aggregation behaviors that vary with the oceanographic environment, resulting in diverse echogram signatures that are challenging to identify with fixed rules. To identify such animals, scientists rely on past experience and other contextual information (e.g., geographic locations, water depth, and season) to create interpretation labels by drawing polygons enclosing these aggregations on echograms (top panel of Fig. 1C). For example, Pacific hake (*Merluccius productus*) a semi-pelagic groundfish that can aggregate both in midwater or near the seafloor and form mixed assemblages with other animals, has been identified mostly through manual annotation (Bailey et al., 1982; Phillips et al., 2023; Godínez-Pérez et al., 2025).

The dramatic increase of echosounder deployment on ocean observing platforms has contributed to a substantial growth of data volume in the past decade. Hundreds of terabytes of echosounder data have been collected from echosounders operating across a wide range of frequencies, signal bandwidths, and sensor configurations (e.g., single-beam, split-beam, multibeam, etc.) mounted on ships, autonomous vehicles, and surface and seafloor moorings (NOAA National Centers for Environmental Information, 2011; Wall et al., 2018; Ocean Observatories Initiative, 2014; Ocean Networks Canada Society, 2020). More flexible automated methods are needed to effectively analyze this new influx of data.

There is a history of using classical machine learning (ML) methods to perform echo analysis, including clustering (Petitgas, 2003; Burgos and Horne, 2008; Campanella and Taylor, 2016), tree-based algorithms (Fernandes, 2009; Fallon et al., 2016; Ji et al., 2020; Proud et al., 2020), and tensor and matrix decomposition (Lee and Staneva, 2019, 2020). More recently, computer vision-based deep learning (DL) models are being rapidly developed to classify echogram features (e.g., Brautaset et al., 2020; Choi et al., 2021; Marques et al., 2021b, 2021a; Ordoñez et al., 2022; Slonimer et al., 2022; Pala et al., 2023; Yassir et al., 2023; Culhane et al., 2025; Handegard et al., 2025; Wibowo et al., 2025; Agarwal et al., 2026). Object detection is one class of such methods, through which a box is drawn around detected scatterers (Marques et al., 2021b). Segmentation is another class of methods that label each pixel on an echogram as a specific taxon or class. These pixel-wise labels are referred to as "masks," and are paired with echograms to train segmentation models (Fig. 1C) (Brautaset et al., 2020; Marques et al., 2021a; Ordoñez et al., 2022; Slonimer et al., 2022; Choi et al., 2023; Vohra et al., 2023). For semantic segmentation, each pixel of an echogram is assigned a class label (e.g., fish species), whereas for instance segmentation, individual instances of the same class (e.g., separate labels for distinct schools of the same fish species) are distinguished (Brautaset et al., 2020; Ordoñez et al., 2022; Choi et al., 2023; Vohra et al., 2023).

Existing literature on DL-based echo analysis focuses on latter stages of echo data preparation, such as concatenating auxiliary non-acoustic data onto the echogram or applying image augmentation methods such as resampling (Ordoñez et al., 2022), or thresholding echo measurements to refine labels (Calin et al., 2026), with little emphasis on how the acoustic data are organized into formats suitable for ML development and analysis in the first place. For example, acoustic-trawl surveys are typically conducted based on linear transects, which are discrete navigational segments designed to enable rigorous statistical analysis of survey observations. Because acoustic data are continuously collected, they must be filtered to remove the off-transect (OT) periods and retain only within-transect (WT) segments (red and black lines, respectively, in Fig. 1B) for downstream analysis. Depending on the target species and survey protocol, human annotators may not label the echograms in a manner suitable for direct usage in ML development. For example, delineating the seafloor is typically a procedure separate from labeling fish aggregations, necessitating additional steps in data organization to remove seafloor echo contamination. In addition, non-acoustic data properties, such as geographic position, may be important in the analysis and should be integrated and propagated downstream in the organized dataset. This consideration is characteristic of echograms and geospatial datasets, such as satellite remote sensing images, where pixel values are intrinsically linked to physical coordinates rather than purely visual content as in generic images.

To address this gap, we present a data organization workflow designed to produce echogram and mask datasets that share the same coordinates and are suitable for semantic segmentation, and provide recommendations for adapting this workflow to generate datasets suitable for other select ML tasks. We also discuss challenges arising from potentially differing spatiotemporal resolutions among acoustic data, annotation labels, and associated environmental datasets. To support transparency and computational scalability, we further provide code and example notebooks that implement the workflow using open-source software packages along with performance benchmarks to demonstrate the capability of the workflow to process large survey datasets.

# Materials and Procedures

In this section, we describe the datasets we use to develop and test this workflow, the workflow details, and our software implementation.

## Example acoustic-trawl survey datasets

Over the past two decades, fisheries scientists from the United States (U.S.) National Oceanic and Atmospheric Administration (NOAA) Northwest Fisheries Science Center (NWFSC) and Canada's Department of Fisheries and Oceans (DFO) conducted the Joint U.S.-Canada Integrated Ecosystem and Pacific Hake Acoustic-Trawl Survey (hereafter referred to as the "Hake Survey") on a biennial basis. Effective management of Pacific hake is critical as they are a key trophic link in the California Current Large Marine Ecosystem and are one of the most abundant commercial fish stocks along the West Coast of the U.S. (Fleischer et al., 2005; Ressler et al., 2007). During 2005-2025, narrowband "power-angle" data (raw echo magnitude and electric split-beam angles) were collected consistently at 18, 38, and 120 kHz, with other transducer channels added depending on the configuration of a specific year. Kongsberg Simrad EK60 and EK80 scientific echosounders were used from 2005-2019 and 2021-2025, respectively. These files are publicly available through the National Centers for Environmental Information (NCEI) Water-Column Sonar Data Collection (NOAA National Centers for Environmental Information, 2011).

The survey typically runs from June to September from Point Conception, California, U.S. to Dixon Entrance north of Haida Gwaii, British Columbia, Canada, with the vessel traversing parallel transects spaced 10-20 nautical miles apart (Fig. 1A). Generally, transects were oriented east-west between the 50-meter and the 1500-meter isobaths, encompassing the continental shelf, shelf break, slope, and edges of the ocean basin. When hake acoustic signature was observed at the end of a transect, the transect was extended until hake acoustic signature was not seen for at least 0.5 nautical mile, to ensure that the full geographic range of the hake stock was surveyed (Fleischer et al., 2005). Concurrent with the acoustic sampling, midwater trawls were conducted to collect biological ground truth data to help confirm the identity of acoustically identified fish aggregations and to provide biometric data such as fish lengths and weights (Simmonds et al., 1992; Fleischer et al., 2005; Thomas et al., 2024).

Conventional echo data analysis was carried out in Echoview (Echoview Software Pty Ltd, 2025) by manually drawing polygons enclosing hake aggregations on echograms (the hake "regions"). As hake aggregate both pelagically and semi-demersally, the lowermost boundaries of the hake regions often trace the seafloor (Fig. 1C). Analysis therefore also included careful delineation of the interface between water column and seafloor and an additional 1m offset was applied to the seafloor delineation to ensure no seafloor echo contamination on near-seafloor hake aggregations occur. In addition, echogram regions containing false bottom (strong seafloor echoes from previous pings that persist in the water column during subsequent pings) or other noise sources were also annotated and removed from further analyses.

Based on these annotations, the echo energy at 38 kHz enclosed within hake-only regions is integrated along depth and used to estimate the abundance and biomass of hake in reference to trawl biological data (Fleischer et al., 2005; Thomas et al., 2024).

## Workflow overview

The data organization workflow described in this paper aims at producing analysis-ready datasets for ML developments. In two stages, the workflow integrates expert annotations with echosounder data files to generate gridded data segments with spatially and temporally aligned echo data and masks that can be used for semantic segmentation. In the first workflow stage focused on transect data organization (Fig. 2A), we parse and convert raw acoustic measurements into physically meaningful quantities using parameters from echosounder calibration, organize the resulting data into segments consistent with transect designations, and optionally regrid the echo data to achieve the desired uniformity along time and depth. In the second workflow stage focused on generating clean fish masks (Fig. 2B), we convert annotations of fish, noise interference, and water column-seafloor interface from the annotation files to pixel-wise "masks" at the same resolution as the organized echo data from the first workflow stage. The masking process is crucial, because the annotations are recorded as polygons or lines delineated by a series of time-depth points, which are different from the discrete time-depth echogram pixel coordinates of the acoustic data. At the end of the two stages, the organized echo data from transect segments and the corresponding masks jointly form an analysis-ready dataset suitable for ML model developments. Under the hood, the workflow uses the Echopype and Echoregions software packages (Lee et al., 2024a, 2024b, 2026b; Tuguinay et al., 2026), which are discussed in the section **Computational implementation**.

## Transect data organization

The first stage of the workflow involves (1) organizing echo data into non-overlapping segments defined by transect breakpoints, and (2) regridding echo measurements in each segment onto a uniform grid along the time-depth coordinates (Fig. 3) conducive to ML developments (uniform image pixels).

The first step of the transect data organization stage is to parse and convert raw acoustic measurements into physically meaningful quantities. For each echosounder data file, calibration coefficients are applied to the raw echo measurements to compensate for transmission loss to obtain physically meaningful volume backscattering strength (Sv; dB re 1 $m^{-1}$). The acoustic data are then partitioned into within-transect (WT) and off-transect (OT) segments (Fig. 3A and 3B). The Hake Survey defines four transect breakpoints: start transect (ST), break transect (BT), resume transect (RT), and end transect (ET). The ST-BT, ST-ET, RT-ET, and RT-BT segments are within-transect (WT) segments (black lines in Fig. 1B) for which the acoustic data are processed for biomass estimation. The BT-RT and ET-ST segments are off-transect (OT) segments not considered in the analysis (red lines in Fig. 1B). OT segments are typically related to transiting between transects or during trawl operations. During data collection, the echosounder systems created files based on a user-specified maximum file size (25 MB from 2005 to 2021 on the EK60 echosounder and 100 MB starting from 2021 on the EK80 echosounder). Therefore, the files generally do not align with the transect breakpoints (Fig. 3B). The OT data sections are trimmed to retain only WT segments. Noise removal can be optionally performed on the Sv data using Echopype [currently supporting background and impulse noise removal (De Robertis and Higginbottom, 2007; Ryan et al., 2015)]. Noise removal is not included here because the Hake Survey echo data contains minimal background and impulse noise, and other noise sources, such as false bottom, are labeled and removed in the workflow Stage 2. Note that some noise removal algorithms, such as impulse noise removal, need to be applied at the raw echo data resolution.

In the second step of this stage, the Sv data segments are optionally regridded into uniform time and depth bins (Fig. 3C). The regridding operation is performed because the echosounder may have operated at variable ping rate, and the range bins in the echo measurement may not be uniform across transducer channels or along depth (when there is a large sound speed gradient). For the Hake Survey datasets where the vessel speed along transect was maintained relatively constant at 10.5 knots, the regridding operation ensures that echograms passed into an ML model have uniform time-depth coordinates to avoid stretching or shrinking the appearance of fish aggregations, which can negatively impact ML model training and performance. If vessel speed varies significantly, such as in the case during slower trawl operations, or when data are collected from wind-powered platforms (e.g., Saildrones), regridding along a geospatial coordinate (e.g., distance traveled) would be more suitable for maintaining fish aggregation morphological features in the echogram images. The regridding operation can be omitted if the uniformity of the original time-depth bins is sufficient for analysis. In addition, regridding should not be performed when the model uses phase information, as phase values cannot be easily interpolated.

In our implementation, the regridding operation is performed by computing the mean volume backscattering strength (MVBS; dB re 1 $m^{-1}$) using 5 s × 1 m bins. There are other methods for regridding, such as “match geometry” (Echoview Software Pty Ltd, 2025) and weighted mean based on echo strength (Met Office, 2013; Ordoñez et al., 2022). We chose to compute MVBS to form a uniform time-depth grid because of its simplicity and wide usage in fisheries acoustics (Madureira et al., 1993; Maclennan, 2002; Fielding et al., 2004; Simmonds and MacLennan, 2005). MVBS regridding is done by computing binned averages of volume backscattering coefficient ($s_V$; $m^{-1}$) measurements enclosed in each regrid box in the linear domain and converted back into the log domain (blue boxes in Fig. 3C and Fig. 3D).

## Clean fish mask creation

The next stage of the workflow converts manually annotated fish regions (polygons) into echogram labels in the form of pixel-wise masks generated in the previous stage (Fig. 4). For supervised ML methods, these masks serve as ground truth labels that assign echogram pixels to different classes according to echogram features, and this assignment is what the model aims to learn during the training phase. The classes in the context of fisheries acoustics are typically different scattering sources, such as specific animal taxa, seafloor echoes, or noise features.

In our implementation, the goal is to create clean hake masks free from any non-fish scattering sources to train a binary (hake vs non-hake) semantic segmentation model. The procedure described below can be easily adapted to generate multi-class masks that include other animal species occurring in the same ecosystem. To emphasize that this approach can be readily-adapted for similar echosounder datasets, hereafter we will use "fish regions" in lieu of hake regions.

In this stage, clean fish masks are created by: (1) converting manual labels in continuous time-depth coordinates to pixel-based masks, (2) removing portions of fish masks that intersect with seafloor masks or masks of any other unwanted echoes, such as those from surface-generated bubble plumes, and (3) regridding the clean fish masks onto the uniform time-depth grid matching the organized echo data generated in transect data organization. Importantly, step 1 needs to be carried out at the same time-depth echogram resolution that the manual annotation was based on before regridding (step 3), to avoid including seafloor echo samples into the fish mask. Due to the effective shift of seafloor location during regridding (Fig. 3D), deviating from this sequence of operations may create erroneous inclusion of seafloor echo samples (*see* section **Supplemental Information Text S1** for details).

Polygon scan conversion, a computational geometry algorithm (Preparata and Shamos, 1985), is used to convert coordinate-based manual labels (polygon defined by a sequence of time-depth vertices) into pixel-based echogram labels (Fig. 4B). Let $(t, d)$ denote a time-depth coordinate of a pixel in the echogram, and let $R_n$ represent the $n$-th annotated fish region (out of $N$ regions). $M_n$ is the $n$-th fish region mask defined pixel-wise using the ray-casting rule:

$M_n(t,d) = $ (number of intersections of a rightward ray with $R_n$ boundary) mod 2

Therefore, if pixel $(t,d)$ lies inside $R_n$, $M_n(t,d) = 1$. If pixel $(t,d)$ is outside of $R_n$, $M_n(t,d) = 0$. The same operation can be performed on the seafloor annotation vertices to create seafloor masks. Note that the ray direction can be chosen arbitrarily.

For fish species aggregating near the seafloor, it is often necessary to clean the fish masks via:

$$M_{n,cleaned}(t,d) = M_n(t,d) \wedge \neg M_{seafloor}(t,d)$$

where $\wedge$ denotes the logical 'AND' operation, and the negation operator $\neg$ inverts the seafloor mask values. As such, a pixel is within the cleaned mask if it is within the original fish mask and outside of the seafloor mask (Fig. 4C).

For semantic segmentation, fish masks within the same WT segment are combined:

$$M_{combined,cleaned}(t,d) = \bigvee_{n=1}^{N} M_{n,cleaned}(t,d)$$

where $\vee$ denotes the logical 'OR' operation. For instance segmentation, where each instance of the same class is distinguished, this mask combination step should be omitted to preserve the separate identity of individual fish aggregations (*see* section **Workflow adaptation for select ML approaches**).

Finally, the clean fish masks are optionally regridded (Fig. 4D) to match the resolution of the organized echo data generated from the transect data organization stage by performing the following operation for each regridding bin containing $N$ points:

$$M'(t',d') = \bigwedge_{(t,d)\ in\ MVBS\ bin}^{N} M_{combined,cleaned}(t,d)$$

## Computational implementation

The workflow is implemented computationally using two open-source Python packages, Echopype (for transect data organization) and Echoregions (for clean fish mask creation). While some workflow steps can be performed using other software, such as pyEcholab (Wall et al., 2018) and Echopy (Ariza et al., 2023), we choose to use Echopype and Echoregions as they are designed to scale with increasing data volumes and computation resources without major code changes. Both packages use Xarray for performing coordinate-aware computations on N-dimensional arrays, with data products following the NetCDF data model that can be serialized in Zarr or NetCDF-4 file formats and stored locally or on the cloud (Lee et al., 2024a, 2024b). In our implementation, we chose Zarr for data serialization because of its cloud-optimized storage architecture. Both Zarr and NetCDF-4 support chunked storage, which divides datasets into smaller, independently accessible blocks. However, NetCDF-4 stores chunks within a single serialized file, whereas Zarr stores them as separate files or objects, which is optimal for cloud-native storage (Pfander et al., 2021; Verma et al., 2023; Martin and Fisher, 2024; Rew et al., 2026). Importantly, Xarray also interfaces with Dask, an open-source Python package, to create Directed Acyclic Graphs (DAGs) to perform out-of-core computations and facilitate “embarrassingly parallel” processing by distributing independent computations across multiple processing units (Rocklin, 2015; Dask Development Team, 2016; Moreno et al., 2022). This allows us to enhance specific functions of Echopype and Echoregions to ensure workflow scalability, which we discuss below. We also recommend that users read the latest Dask documentation for general best practices (Dask Development Team, 2016).

The most computationally expensive step in the transect data organization stage is to regrid the organized echo data by computing MVBS, due to the required grouping of echo data samples that are contained within specific time-depth bins (the blue "regrid boxes" in Fig. 3C). We enhanced the Echopype `compute_MVBS` function using Flox, a Python package integrated with Xarray that improves the efficiency of grouped aggregations on Dask arrays (Cherian, 2023). Without Flox, Xarray defaults to grouping with the "split-apply-combine" method, through which the data samples are split into groups, the desired function (e.g., mean) is applied on each of these groups, and the group outputs are combined back into a single dataset. This approach does not perform well with large data volume, as the memory and computation cost increases with the number of groups. Flox circumvents this limitation by analyzing the dataset to understand how data samples are positioned with respect to their groupmates, and making use of these relationships in an efficient vectorized algorithm to perform the grouping operation. With the Flox integration, users can use `compute_MVBS` to run on arbitrarily large acoustic datasets using regular personal laptop specifications (e.g., 4 CPUs and 16 GB of RAM).

The most computationally expensive step in the clean fish mask creation stage is to perform polygon scan conversion to create fish masks (Fig. 4B). This is performed using the Echoregions `Regions2D.region_mask` method, in which we accelerated the mask creation by wrapping functions from Regionmask, a Python package to create geospatial masks, with Dask to enable parallelization. Under the hood, Regionmask uses Shapely, a Python package for geometric operations, which in turn uses the Geometry Engine Open Source (GEOS), a C++ package for computational geometry, to perform the polygon scan conversion operation (Gillies et al., 2024; Hauser et al., 2024; GEOS contributors, 2025).

We provide minimally reproducible examples of the workflow in the form of executable Jupyter notebooks containing interactive Python code, visualizations, and corresponding text explanations (Lee et al., 2026a). The first notebook,

`single_transect_ML_data_organization_with_regridding.ipynb`, demonstrates the complete workflow with the regridding steps included. The second notebook, `single_transect_ML_data_organization_without_regridding.ipynb`, demonstrates the workflow without regridding. The notebooks start with example EK80 echosounder `.raw` data files and Echoview-exported region annotation `.evr` and seafloor delineation `.evl` files. Echopype and Echoregions convert these proprietary formatted files into interoperable formats native to the scientific Python ecosystem, enabling further processing and analysis without the need for proprietary software. Alongside the workflow-specific steps, the notebooks also describe optional echo data processing steps (e.g., rule-based noise removal), and how to modify the masks into forms suited for select ML approaches *(see* section **Workflow adaptation for select ML approaches**). The workflow can be adapted for acoustic and annotation datasets in `.zarr` and tabular formats supported by Echopype (e.g., `.01A` files from AZFP echosounders) and Echoregions (e.g., `.csv` files).

# Assessment

## Computation scalability

We benchmarked the computational scalability of our workflow implementation on the U.S. sections of the 2019 and 2023 Hake Survey data (NOAA Northwest Fisheries Science Center, 2022, 2025). The 2019 and 2023 acoustic datasets were collected with an EK60 and an EK80 echosounder, respectively. Given the same echosounder configuration for ping rate (determined by the "Maximum Ping Mode" in the configuration software; typically <= 4 s) and recording range (750 m), EK80 samples the echo time series at a much higher resolution along range than EK60, resulting in a large difference of the total data volume (Table I). The annotation files are small (<100 KB) and do not contribute significantly to the total data volume. However, the number of regions annotated within each transect segment does impact the computing speed (see below).

"Wall clock time" was selected as the metric to evaluate computational efficiency (Table I and Fig. 5) because it provides a comprehensive measure of computational performance: it captures the total elapsed time experienced by the user, including computational overhead associated with Dask task scheduling and execution, as well as the I/O operations for Zarr-serialized files. It is also the best metric for estimating costs associated with cloud computing services (e.g., Amazon Web Services Elastic Compute Cloud, Google Compute Engine), where billing is based on the virtual instance wall clock time.

In general, the total and per-transect segment wall clock time increases with increasing data volume or decreasing computing resources (Table I and Fig. 5). Specifically, in Fig. 5, the slope of the fitted regression line represents the general trend of computational speed: the lower the slope, the shorter the wall clock time given the same size of data. Importantly, for the clean fish mask creation stage (Fig. 5B and Fig. 5D), the transect segments with wall clock times far above their respective regression line contain many more fish regions than those near the regression line. The benchmarking results also show that the computational efficiency gain plateaus for the transect data organization stage with larger resources, whereas the efficiency continues to improve for the clean fish mask creation stage (Fig. 5, comparing the change from the magenta to red lines between the left and the right panels). This can be understood by observing the associated Dask graphs that represent computational tasks as nodes and their interdependencies as edges. In general, larger Dask graphs introduce more complex scheduling and larger communication overhead across workers, which negatively impacts the efficiency of parallel computation. The computations in the transect data organization stage involve deeper Dask graphs, suggesting that many tasks cannot run in parallel and depend on outputs from upstream tasks. Therefore, increasing the number of CPUs and thus the number of workers capable of executing parallel tasks may not necessarily shorten the wall clock time for this stage.

## Workflow adaptation for select ML approaches

The workflow presented here is designed to organize data for semantic segmentation tasks but can be adapted to organize data for select ML tasks with only small modifications. For example, to create data for instance segmentation that requires each instance of the same class to be separately labeled (Marques et al., 2021a; Slonimer et al., 2022), the step to combine individual fish masks in the clean fish mask creation stage can be omitted to preserve the identity of different fish region annotations in the resulting masks (Fig. 6A). To create data for object detection that requires a bounding box for each object (Marques et al., 2021b), the four vertices of each rectangular box can be calculated from each time-depth polygon annotation or each fish mask (Fig. 6A) and merged for multi-object detection (Fig. 6C). In addition, the WT echo data products from the transect data organization stage can be directly leveraged for unsupervised ML methods such as clustering and autoencoders (Petitgas, 2003; Burgos and Horne, 2008; Campanella and Taylor, 2016; Culhane et al., 2025), tree-based algorithms (Fernandes, 2009; Fallon et al., 2016; Ji et al., 2020; Proud et al., 2020), and tensor and matrix decomposition (Lee and Staneva, 2019, 2020).

# Discussion

In this paper, we present a two-stage workflow to generate analysis-ready fisheries acoustic datasets to support ML development and applications. We detail the operations and our computational implementation in each workflow stage and evaluate the performance on two example fisheries acoustic-trawl datasets. We also provide minimally reproducible executable notebooks that demonstrate the workflow, and discuss modifications required to adapt the workflow to organize data for select ML approaches. Below we discuss additional considerations in applying and extending this workflow, and highlight the need for creating benchmark datasets in fisheries acoustics to accelerate data science and ML development in this field.

The transect data organization workflow stage provides a framework to partition echosounder data into conceptually meaningful segments according to the data collection scenario. For the two example datasets, the acoustic data are partitioned into interleaving WT and OT segments (Fig. 4B). For surveys that employ a zig-zag traversing pattern (Overholtz et al., 2006; Lian et al., 2022), the data can be partitioned based on the turning points, or corners, of the ship track. For mooring deployments at fixed locations (De Robertis et al., 2018; Lee and Staneva, 2020), the data can be partitioned on a daily, weekly, or monthly basis depending on the exact sampling scheme. For glider and profiler-based deployments (Suberg et al., 2014; Benoit-Bird et al., 2018; Grassian et al., 2023), the data can be partitioned according to the down-up dive cycles. In addition to post-processing, this data partitioning framework can also be employed in real-time data collection scenarios if temporal or spatial “breakpoints” can be clearly defined through either manual annotation or automatic methods.

The clean fish mask creation stage assumes that the annotations are sufficient to create clean fish masks and does not itself improve the annotation quality. For the example datasets, the two primary assumptions are that some fish region annotations may overlap with the seafloor, and that the seafloor annotations encompass the entire WT echo data segment. Consequently, any portions of the fish region annotations extending below the seafloor will be removed during clean fish mask creation. If seafloor annotations are unavailable, there exist automated seafloor detection methods that can be applied on echograms to extract the seafloor boundary (Trenkel et al., 2009; Ladroit et al., 2020; Echoview Software Pty Ltd, 2025). Fish and seafloor annotation can be potentially improved using other methodology (Calin et al., 2026; Possenti et al., 2026).

A key property of the presented workflow is that every pixel of the resulting echogram images and corresponding masks is strictly aligned with a physically meaningful coordinate, specifically transducer channel (`channel`), ping transmission time (`ping_time`), and water column depth (`depth`) (see Fig. 2). This explicit coordinate alignment enables the propagation of ancillary data, such as geospatial locations (e.g., latitude and longitude) and other environmental measurements (e.g., salinity, temperature), alongside the acoustic data throughout the workflow, preserving contextual information that can be important for ML model inference and data interpretation. For example, water column attributes such as depth may provide useful information for ML decision-making (Ordoñez et al., 2022). Environmental context, such as spatial relationships with eddies and submarine canyons, is known to influence distribution patterns for hake and important prey species such as euphausiids (Santora et al., 2018; Phillips et al., 2023; Derville et al., 2025; Díaz-Astudillo et al., 2025). Further downstream, many biological quantities are estimated as a function of geospatial locations, in particular when the estimation procedure requires incorporating data from other sources, such as trawl catches (Thomas et al., 2024).

The workflow presented here establishes a structured and computationally scalable pathway to organize fisheries acoustic data for ML developments. It is an important step toward developing benchmark datasets that can be used to evaluate and compare ML models and approaches in this field. Such benchmark datasets have proven instrumental in driving rapid advances in computer vision [e.g., ImageNet (Deng et al., 2009) and COCO (Lin et al., 2014)], and have also recently been developed in passive acoustic monitoring (PAM), a field closely related to fisheries acoustics, to facilitate development of automatic animal call detection and sound event classification algorithms [e.g., DeepShip (Irfan et al., 2021) and animal sound datasets (Hagiwara et al., 2023; Palmer et al., 2025)]. Specifically, we note that for benchmark datasets, it would be advantageous to provide both the consolidated, analysis-ready datasets (outputs of the presented workflow) and the corresponding raw sensor data with annotations (inputs of the presented workflow), along with example preprocessing workflow code. The former can be used directly to develop and evaluate models that rely on data features already available in the consolidated datasets (e.g., Sv echograms for models treating echo data as images), whereas the latter enables reprocessing of the sensor data to extract additional features (e.g., phase information from echo pressure waveforms).

Looking ahead, we view the workflow and the accompanying software implementation presented here not as a prescriptive standard, but as a starting point for discussion, experimentation, and collaboration around datasets and annotations for ML development in the fisheries acoustics community. We envision community-wide efforts to collaboratively refine these procedures and tools, define data sharing conventions, and develop practical mechanisms for data sharing, which will help support transparency, reproducibility, and meaningful comparison across models. Such coordinated efforts would lower barriers to entry for ML research, accelerate methodological progress, and help ensure that these advances in automated echo data analysis are robust, interpretable, and broadly transferable.

# Conflict of interest

The authors declare no conflicts of interest to disclose.

# Author contribution statement

The author contributions are as follows: Wu-Jung Lee and Valentina Staneva conceived the workflow and developed the initial codebase. Caesar Tuguinay developed the scalable implementation, tested it on multiple datasets, and evaluated its performance together with Wu-Jung Lee. Elizabeth M. Phillips, Rebecca E. Thomas, Alicia Billings, and Julia Clemons collected and annotated the echosounder data and contributed contextual knowledge that informed workflow development. Caesar Tuguinay and Wu-Jung Lee wrote the manuscript, and all other authors reviewed and edited the manuscript.

# Data availability statement

The ship data were collected as part of the 2019 and 2023 Joint U.S. and Canada Pacific Hake Integrated Acoustic and Trawl Survey (NOAA Northwest Fisheries Science Center, 2022, 2025) and are available in the NOAA NCEI Water-Column Sonar Data Archive Amazon Web Service S3 bucket: registry.opendata.aws/ncei-wcsd-archive/. The full set of annotations used in the computational benchmark example are available from the NOAA NWFSC Fisheries Engineering and Acoustic Technologies (FEAT) team upon reasonable request.

The latest releases of the software packages or repositories have been archived with Zenodo: Echopype (doi:10.5281/zenodo.3906999), Echoregions (doi:10.5281/zenodo.8400849), and Echopype-examples (doi:10.5281/zenodo.5618177).

# Acknowledgements

We thank Steve de Blois, John Pohl, and Dezhang Chu from the NOAA Northwest Fisheries Science Center for their effort in data collection and echogram annotation. This work used computational and storage resources from Jetstream2 at Indiana University and the Open Storage Network through allocations CIS240371 and AGR230002 from the Advanced Cyberinfrastructure Coordination Ecosystem: Services & Support (ACCESS) program, which is supported by U.S. National Science Foundation grants #2138259, #2138286, #2138307, #2137603, and #2138296. This work was supported by NOAA grants #NA20OAR4320271 and

#NA24NMFX405C0016 awarded through the Cooperative Institute for Climate, Ocean, and Ecosystem Studies (CICOES), University of Washington.

# Figures

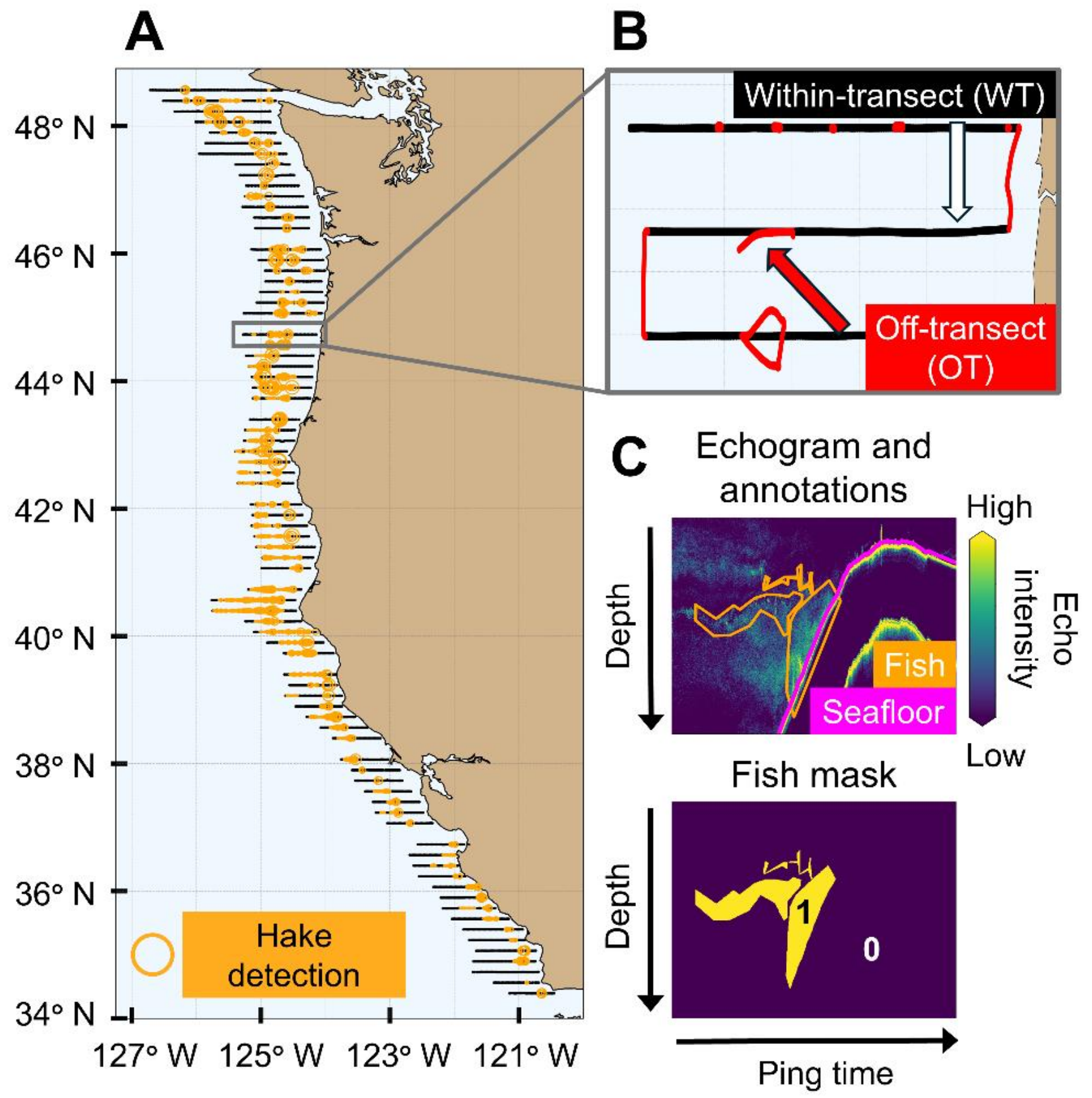

**Fig 1.** Example data from the NOAA Northwest Fisheries Science Center's Pacific Hake Survey along the West Coast of the United States, a typical fisheries acoustic-trawl survey. (A) Survey transects and the distribution of hake detection. The size of the orange circles scales with the acoustically inferred hake abundance. (B) Zoomed-in section showing the within-transect (WT) and off-transect (OT) segments of the ship track. Only the WT segments are used for hake abundance estimation. (C) Top: Echogram and annotations. Bottom: binary masks indicating echogram pixels containing hake (fish; mask value = 1) and non-hake (mask value = 0).

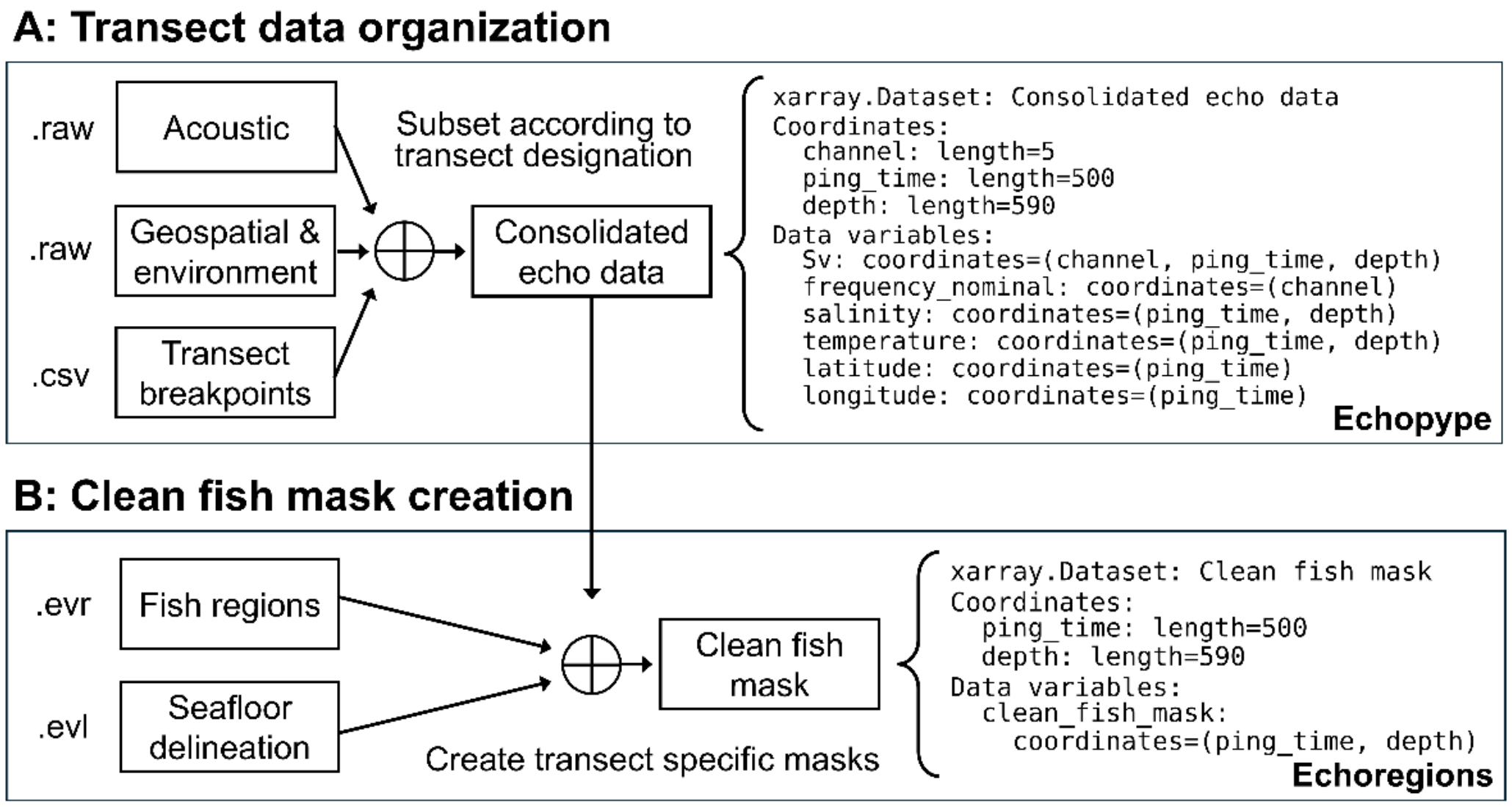

**Fig 2.** Workflow overview. (A) Transect data organization: Acoustic data, ancillary data (e.g., geospatial locations and environmental variables), and transect breakpoints (e.g., start and end times) are consolidated into standardized transect datasets stored as individual `.zarr` files. The acoustic and ancillary data are typically raw instrument-generated data files, such as Simrad echosounder `.raw` files. The transect breakpoints are typically recorded in tabular format, such as `.csv` files. (B) Clean fish mask creation: Fish region and seafloor annotations are combined with the consolidated echo data to create clean fish masks free from non-fish scattering sources. Annotations may be produced manually or automatically. In our example, Echoview `.evr` and `.evl` files define hake aggregations and the water column-seafloor interface, respectively. The workflow uses the open-source Echopype and Echoregions package to generate organized datasets aligned along the same set of coordinates.

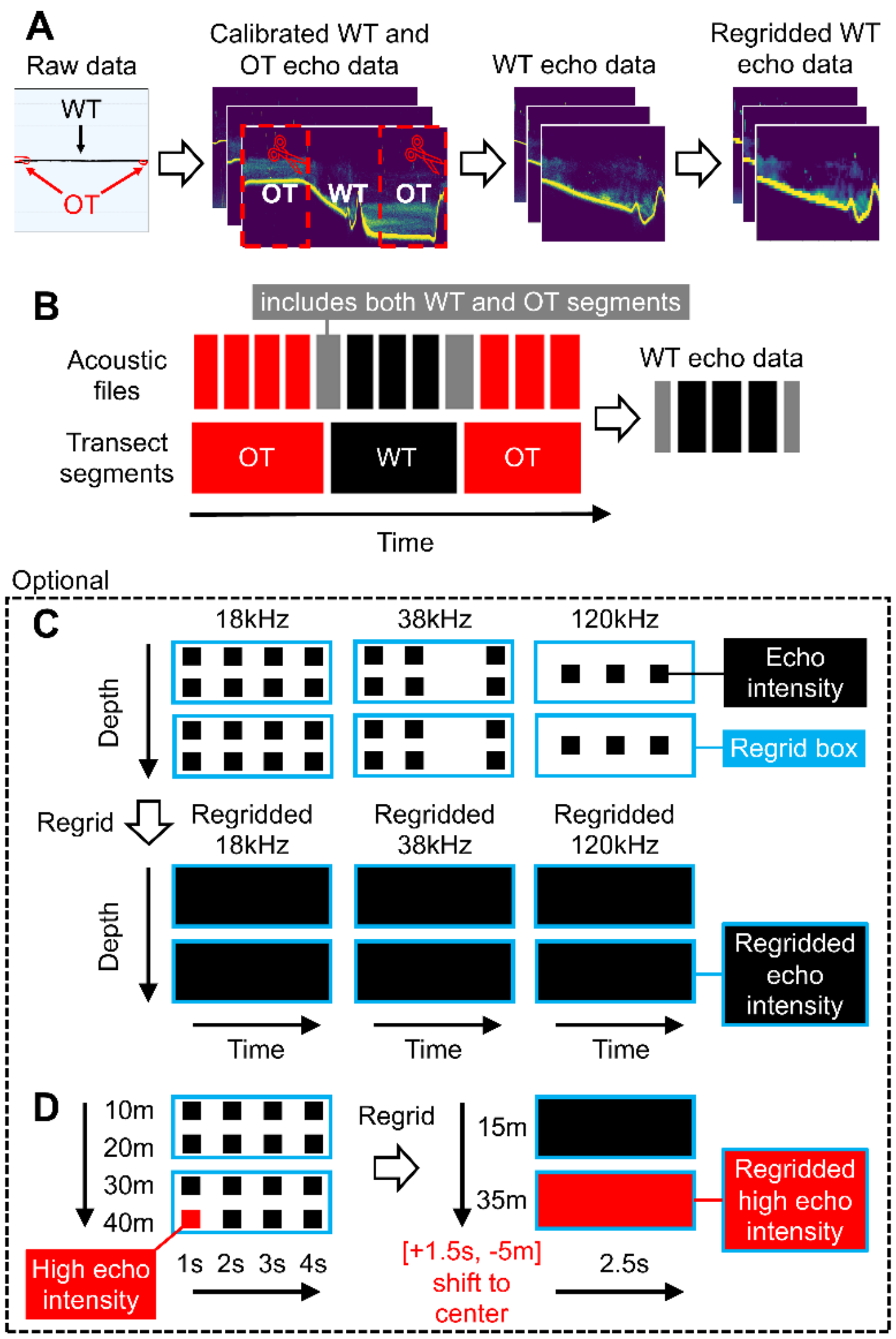

**Fig 3.** Summary of the steps involved in the transect data organization workflow stage. Detail procedures are illustrated in panels B-D. (B) WT echo data is generated by combining all echo data intersecting with the WT time points and trimming off any OT sections. (C) WT echo data are regridded by computing MVBS for each regrid bin, since the original echo measurements may not be uniformly spaced along the time and depth coordinates (see text for detail). (D) An illustration of how the regridding operation creates an effective shift of the time-depth coordinate of the echo data. In panels C-D, each square represents an echo intensity measurement (an echogram pixel at the raw resolution). Note that the regridding operation is optional (*see* section **Transect data organization** for details).

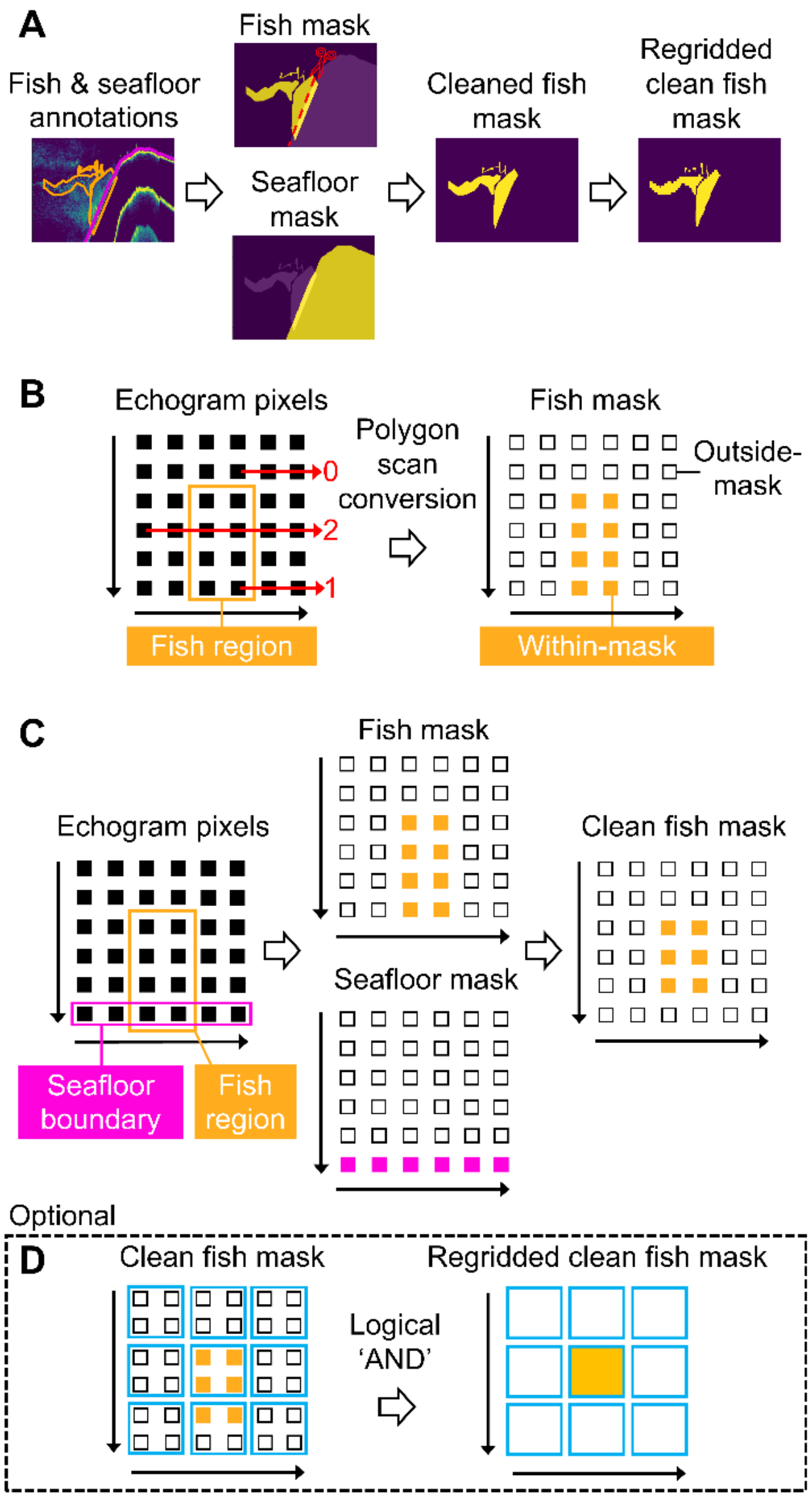

**Fig 4.** (A) Summary of the steps involved in the clean fish mask creation workflow stage. Detailed procedures are illustrated in panels B-D. (B) Illustration of polygon scan conversion: pixels with an odd or an even number of ray–boundary intersections are inside and outside the polygon, respectively. (C) Echogram pixels within a fish mask that are also covered by seafloor masks are removed to create a clean fish mask. (D) The clean fish mask is regridded to match the resolution of the organized echo data from transect data organization. Note that the regridding operation is optional (*see* section **Transect data organization** for details).

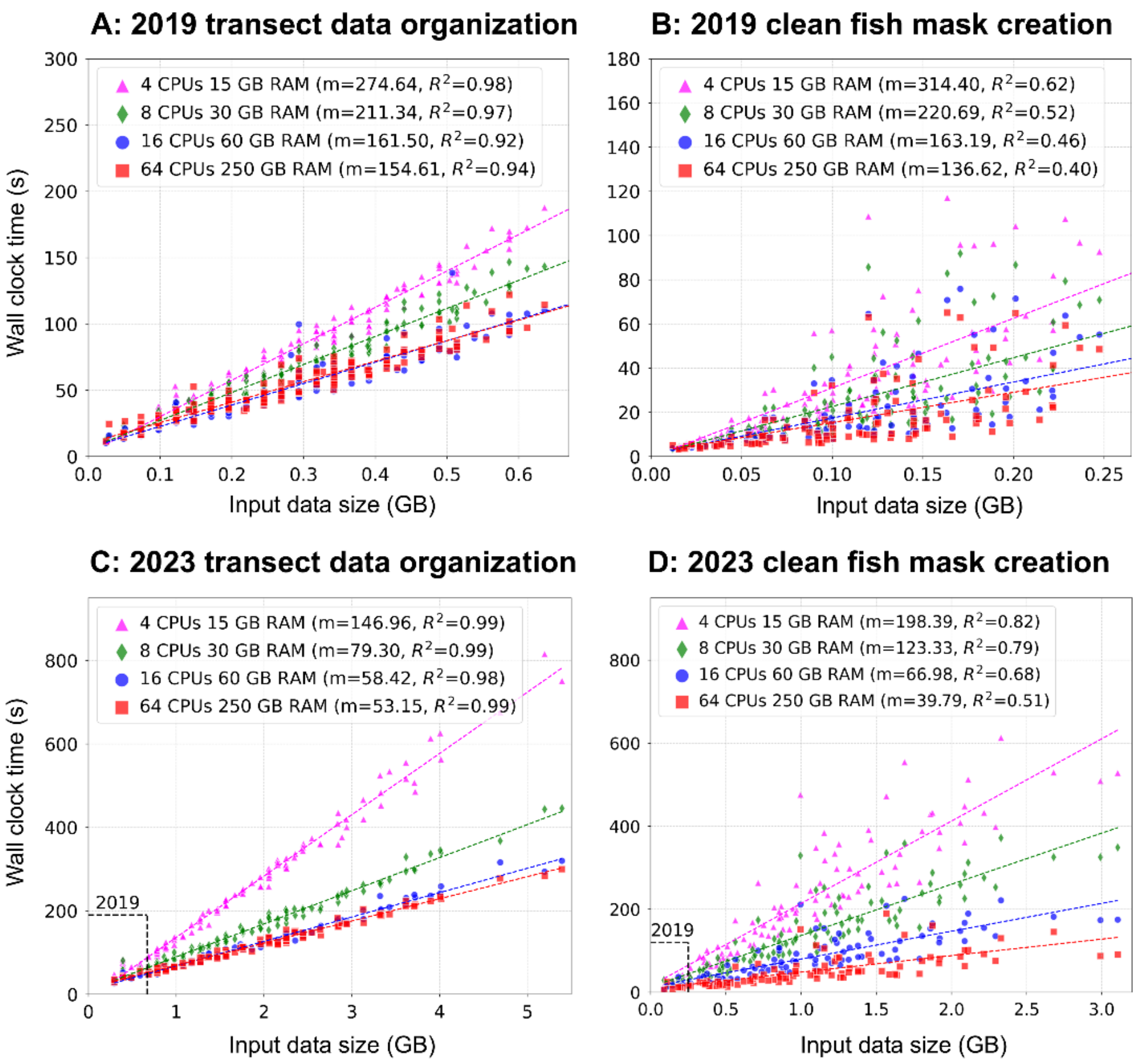

**Fig 5.** Distributions of wall clock time for 2019 (A-B) and 2023 (C-D) WT segments across input data size given different levels of computing resources. Only WT segments containing both fish and seafloor annotations are included, as these segments require all processing steps within each workflow stage. Dashed lines are the best fit linear regression given a computing resource level with slope (m) and coefficient of determination ($R^2$). In general, the larger the computing resources, the smaller the slope, indicating a decrease in total wall clock time. The box in the lower-left corners of panels C-D indicate the extent of panels A-B, since the data volume is much smaller for 2019 than for 2023. An outlier 2019 WT segment with input data size 1.3 GB (for transect data organization) and 0.5 GB (for clean fish mask creation) is omitted in the regression calculation and the plots.

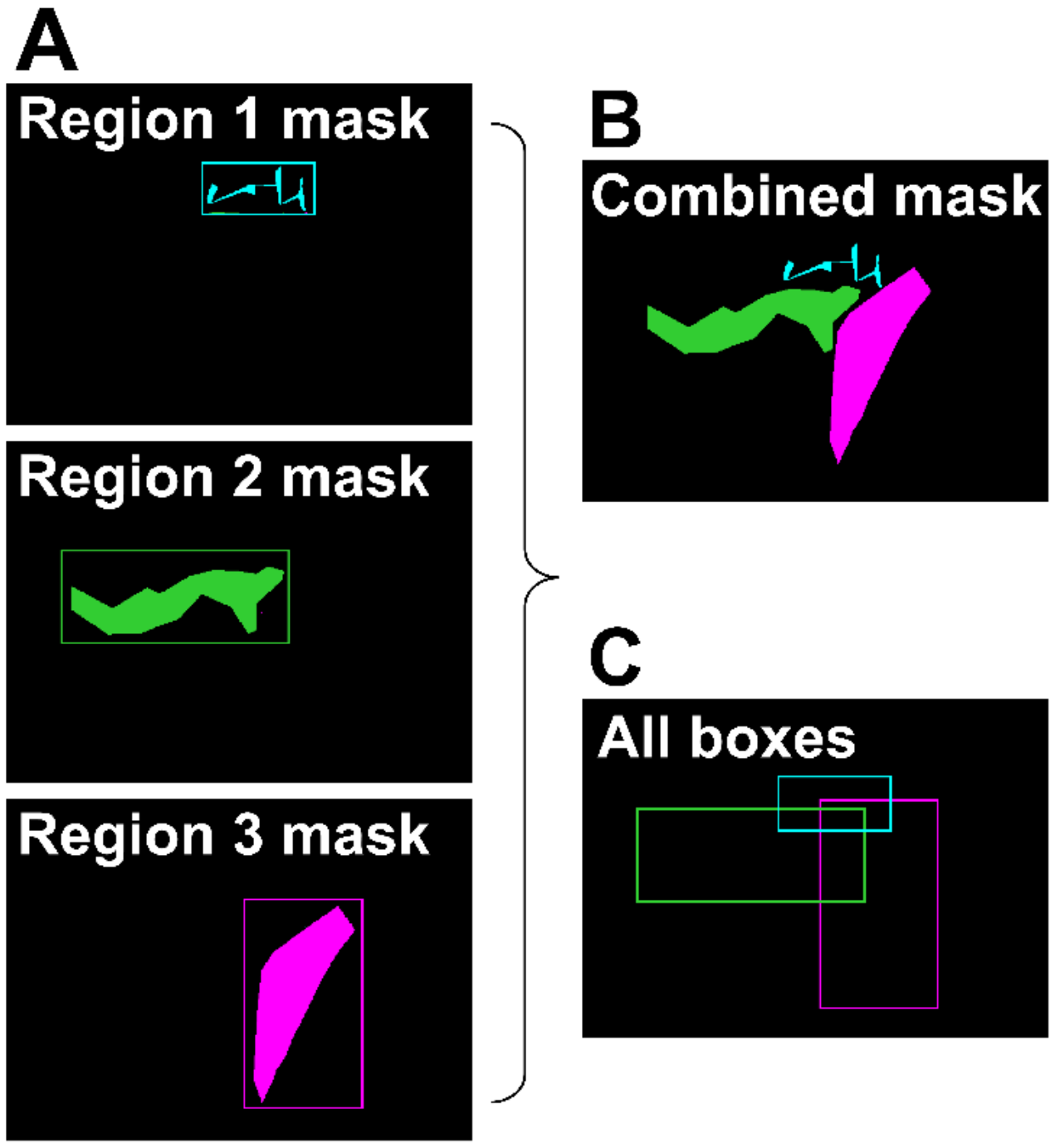

**Fig 6.** Illustration of workflow variations to generate data for select ML tasks. (A) Separate individual masks for three distinct fish regions that are used for instance segmentation. (B) A combined mask for semantic segmentation created by merging the individual region masks in panel A. (C) Multiple bounding boxes for object detection derived from individual masks. Each bounding box is defined by the extents of an individual mask in panel A.

# Tables

**Table I.** Total data volume and wall clock time across computing resources for the 2019 and 2023 data.

| **Year** | **Echosounder system** | **Total acoustic data size** | **Number of files** | **4 CPUs, 15 GB RAM** | **8 CPUs, 30 GB RAM** | **16 CPUs, 60 GB RAM** | **64 CPUs, 250 GB RAM** |
|---|---|---|---|---|---|---|---|
| **2019** | EK60 | 120.9 GB | 5020 | 17345.7 s | 13846.5 s | 10944.9 s | 10928.6 s |
| **2023** | EK80 | 868.8 GB | 8971 | 76697.6 s | 47887.6 s | 32834.6 s | 27759.3 s |

# Supplemental Information

## Text S1 - Sequence of operations to create clean fish masks

In the **Clean fish mask creation** section, we present a sequence of operations to create clean fish masks that match the resolution of the organized echo data generated from the transect data organization stage: the fish and seafloor masks are created at the original echo data resolution, and the fish masks are cleaned by removing overlapping seafloor mask pixels before the fish masks are regridded (Fig. S1A). We analyzed other possible sequences of operations and found that exact order of mask creation, mask cleaning, and regridding is crucial to ensure seafloor echoes are correctly removed from the final fish masks. Below we discuss three other possible sequences of operations, from which only one correctly produces the clean fish mask at the regridded resolution:

1. Fish and seafloor masks are created at the regridded resolution before the fish mask is cleaned. Due to the shift of the echo data coordinates introduced by regridding (see Fig. 3D), the masking operation may miss the seafloor annotation, resulting in seafloor echoes not correctly removed in the final fish mask (Fig. S1B).
2. Fish and seafloor masks are created at the original resolution, with logical 'AND' and logical 'OR', respectively, used during regridding. The resulting empty seafloor mask causes the seafloor echoes not correctly removed in the final fish mask (Fig. S1C).
3. Fish and seafloor masks are created at the original resolution, with logical 'AND' used during regridding (Fig. S1D). This sequence of operations generates the equivalent correct clean fish mask as the sequence used in the workflow presented in the main text (Fig. S1A).

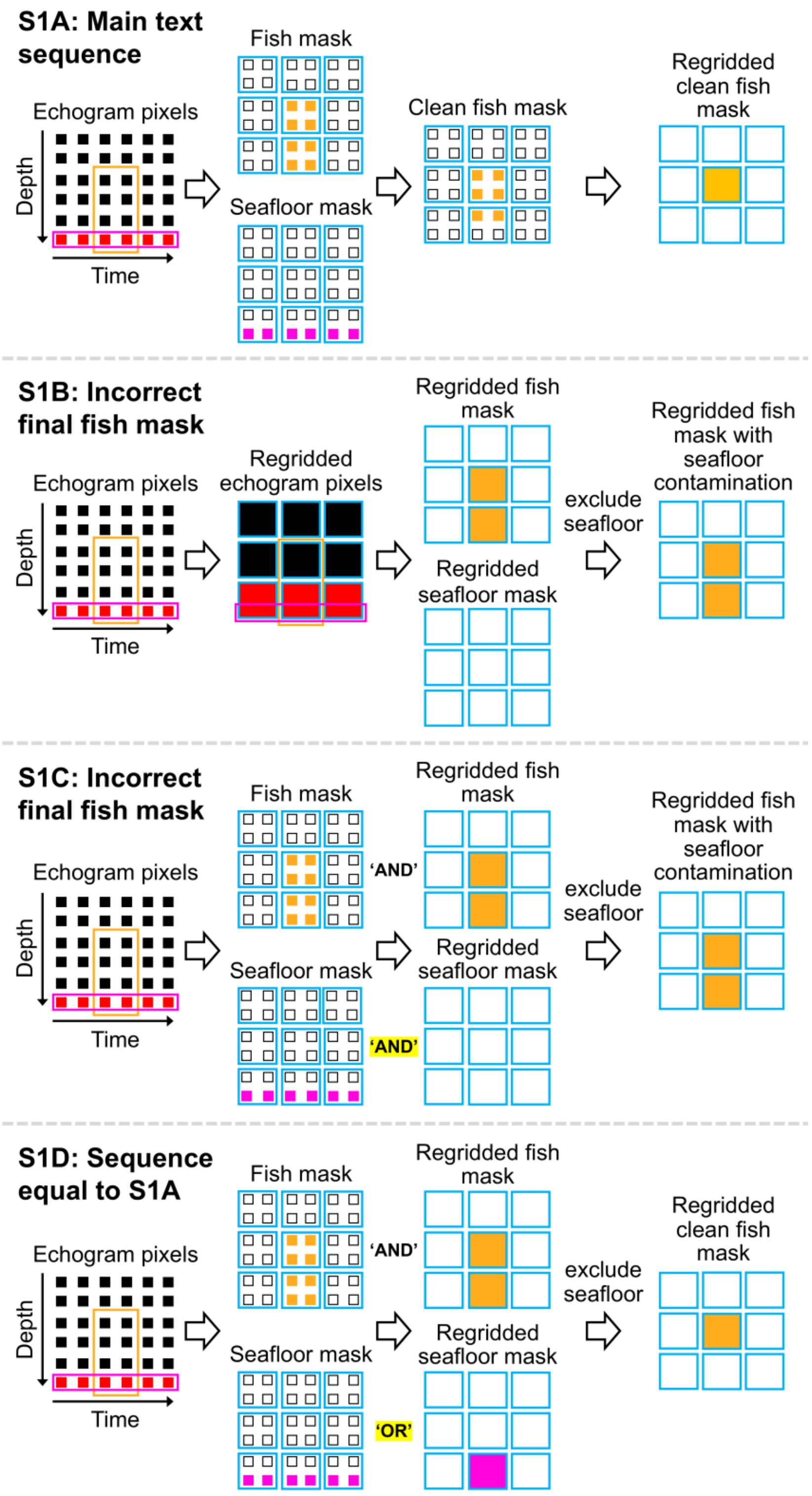

**Fig S1.** Illustration of four possible sequences of operations attempting to generate clean fish masks at the regridded resolution. Note only the sequence illustrated in panel D correctly reproduces the clean fish mask as in panel A. *See* **Supplemental Information Text S1** for detailed descriptions.